%% file: main.tex
\documentclass[conference,peer-reviewed]{aesconf}

\graphicspath{{./}{figures/}}

\usepackage[utf8]{inputenc}

\usepackage{microtype}

\usepackage[numbers,square]{natbib}

\usepackage{booktabs}
\usepackage[dvipsnames]{color}
\usepackage{url}

\usepackage{graphicx}
\usepackage{amsmath}
\usepackage{amsfonts}
\usepackage{amssymb}
\usepackage{hyperref}
\usepackage{siunitx}
\usepackage{import}
\usepackage{tabularx}
\usepackage{enumitem}
\usepackage{quoting}
\quotingsetup{vskip=1pt}

\usepackage{pgfplots}
\usepgfplotslibrary{units, groupplots}
\pgfplotsset{unit markings=parenthesis, compat=1.18}

\makeatletter
 \DeclareRobustCommand\ref{
    \@ifstar\@refstar\T@ref
  }
  \DeclareRobustCommand\pageref{
    \@ifstar\@pagerefstar\T@pageref
  }
 \makeatother

\newcommand{\R}{\mathbb{R}}

\newcommand{\y}{\mathbf{y}}
\newcommand{\x}{\mathbf{x}}
\newcommand{\haty}{\hat{\mathbf{y}}}

\newcommand{\vv}{\mathbf{v}}

\title{Sound Matching an Analogue Levelling Amplifier Using the Newton-Raphson Method}

\author[1]{Chin-Yun Yu}
\author[1]{Gy\"orgy Fazekas}

\affil[1]{Centre for Digital Music, Queen Mary University of London, London, UK}

\correspondence{Chin-Yun Yu}{chin-yun.yu@qmul.ac.uk}

\lastnames{Yu and Fazekas}

\shorttitle{Analogue Levelling Amplifier Emulation}

\begin{document}

\twocolumn[
    \maketitle 

    \begin{onecolabstract}
        Automatic differentiation through digital signal processing algorithms for virtual analogue modelling has recently gained popularity.
        These algorithms are typically more computationally efficient than black-box neural networks that rely on dense matrix multiplications.
        Due to their differentiable nature, they can be integrated with neural networks and jointly trained using gradient descent algorithms, resulting in more efficient systems.
        Furthermore, signal processing algorithms have significantly fewer parameters than neural networks, allowing the application of the Newton-Raphson method.
        This method offers faster and more robust convergence than gradient descent at the cost of quadratic storage.
        This paper presents a method to emulate analogue levelling amplifiers using a feed-forward digital compressor with parameters optimised via the Newton-Raphson method.
        We demonstrate that a digital compressor can successfully approximate the behaviour of our target unit, the Teletronix LA-2A.
        Different strategies for computing the Hessian matrix are benchmarked.
        We leverage parallel algorithms for recursive filters to achieve efficient training on modern GPUs.
        The resulting model is made into a VST plugin and is open-sourced.
    \end{onecolabstract}
]

\section{Introduction}

Controlling the dynamic range of an audio signal is a fundamental task in audio engineering.
Audio effects such as compressors, limiters, and expanders are designed to achieve this goal and are often used in music production, broadcasting, and live sound reinforcement.
Analogue compressors developed in the early days have iconic sounds, such as the popular Teletronix LA-2A, which is known for its smooth and musical compression.
Considerable interest has been devoted to replicating their characteristics in the digital world, known as virtual analogue (VA) modelling~\citep{eichas2017virtual}.

\subsection{Problem Definition}
\label{ssec:problem_def}
Let us denote an audio stimulus as $x(t)$ and its discrete version as $\x = [x(0), x(T), \ldots, x((N-1)T)]^\top \in \R^N$ with $T \in \R_+$ being the sampling period.
We want to model a target analogue compressor $\delta: \R^\infty \rightarrow \R$, specifically, its behaviour to the stimulus $x(t)$ as $y(t) = \delta(x(s) \mid s \leq t)$.
We denote $\y = [y(0), y(T), \ldots, y((N-1)T)]^\top \in \R^N$ as the sampled compressed signal.
Given a controllable base model $f: \R^N \times \R^M \rightarrow \R^N$ parametrised by $M$ parameters, we aim to find the optimal parameters $\theta^* \in \R^M$ that minimises the distance between the output $\hat{\y} = f(\x, \theta)$ and $\y$ according to a distance function $\mathcal{D}: \R^N \times \R^N \rightarrow \R_+$.

\subsection{Related Work}
\label{ssec:related_work}

In recent years, neural networks (NNs) with various architectures have become popular base models for $f$.
\citet{hawley_signaltrain_2019} are pioneers in this approach, using an U-Net-like autoencoder with skip connections to model the LA-2A end-to-end.
\citet{steinmetz_efficient_2022} proposed temporal convolutional networks (TCNs) to enhance efficiency while maintaining large receptive fields of the networks.
Different conditioning strategies have been explored to improve the modelling ability~\citep{comunita_modelling_2023, yeh_hyper_2024}.
More recent works explored the use of state-space models for VA modelling~\citep{simionato2024comparative, simionato2025modeling}.
The choice of architecture is crucial since compressors are used in real-time applications, and the computational cost is a concern.
Despite impressive results, the black-box nature of NNs makes the learnt parameters difficult to interpret and limits creative control.

Among previous works, \citet{wright2022grey} proposed a grey-box approach that combines NNs and signal processing components.
A feed-forward compressor is used as $f$ with a lightweight Gated Recurrent Unit (GRU) at the end to model time-varying non-linear make-up gain.
This approach achieves competitive results while requiring only 5\% of arithmetic operations compared to a single GRU.
The learnt control parameters also give us a better understanding of the target unit's behaviour.
\citet{ycy2024diffapf} further improve their training speed by implementing a specialised kernel for backpropagating the gradients through the ballistics module of the compressor.

\subsection{Motivation}
\label{ssec:motivation}

Both~\citep{wright2022grey} and~\citep{ycy2024diffapf} use gradient descent optimisation to find the optimal feed-forward compressor parameters for VA modelling.
Gradient descent is widely used in training neural networks instead of more robust methods like Newton-Raphson (NR) due to its simplicity and lower storage cost, favouring methods with many parameters ($M \gg 10^6$).
However, since the feed-forward compressor has only a handful of parameters ($M < 10$), the quadratic storage cost of NR is feasible on modern computers.
NR converges faster and more robustly than gradient descent when the objective function is convex around the solution and twice differentiable.
This motivates us to explore the NR method for VA modelling of analogue compressors.
Although NR has been used to solve differential equations in white-box models, such as wave digital filters~\citep{alberto2021wdf} and the physical modelling of musical instruments~\citep{Bilbao2015}, to the authors' knowledge, it has not been used for grey-box modelling of analogue compressors.

\subsection{Contributions}
\label{ssec:contrib}

In this article, we formulate VA modelling as a sound matching problem, aiming to find the optimal parameters of a digital compressor $f$ with the closest match to the target sound.
We verify the feasibility of using the NR method for modelling the LA-2A compressor.
Moreover, we further accelerate the backpropagation method from~\citep{ycy2024diffapf} on modern GPUs using parallel algorithms for recursive filters.
The learnt mapping from the circuit's peak reduction to the parameters is presented, providing an interpretable and intuitive way to control the compressor.
We made the resulting model into a VST plugin, open-sourced under the MPL-2.0 license\footnote {\href{https://github.com/aim-qmul/4a2a}{github.com/aim-qmul/4a2a}}, for validation and as a viable music production tool.

\section{Methodology}
\label{sec:method}

Continuing the work from~\citep{wright2022grey, ycy2024diffapf}, we use a feed-forward compressor as the base model $f$ for its efficiency and high modelling capability.
We will discuss the choice of distance function in Section~\ref{ssec:newton}.
Since the input signal $\x$ is fixed during optimisation, we use $f_\x: \R^M \rightarrow \R^N$ to denote the curried $f$ given $\x$, to simplify the later derivative notations.

\subsection{Feed-Forward Compressor ($f$)}
\label{ssec:ff}
We adopt \texttt{torchcomp}~\citep{ycy2024diffapf}, a differentiable implementation of the feed-forward compressor from~\citep{dafx_comp}.
A digital compressor is defined as:
\begin{equation}
    \hat{y}[n] = x[n]g[n] = x[n] g\left(s[m], \theta \mid m \leq n\right),
\end{equation}
where $g: \R^{n+1} \times \R \rightarrow \R_+$ is the gain reduction function with parameters $\theta$ and $s[m]$ is the side-chain signal.
The design is feed-forward when $s[m] = x[m]$~\citep{giannoulis_digital_2012}.
We remove the root mean square (RMS) level detector from the design for two reasons.
Firstly, Table 7 in~\citep{ycy2024diffapf} shows the converged RMS smoothing coefficient is close to one, implying the smoothing effect is negligible.
Secondly, we found that when running the NR method, the converged $\mathcal{D}(\hat{\y}, \y)$ is slightly lower without it.
A possible explanation is that the extra degree of freedom introduced by the RMS filter creates a local minimum that traps the NR optimisation, a phenomenon we will discuss in Section~\ref{ssec:newton}.

\begin{figure}
    \centering
    \includegraphics[width=\columnwidth]{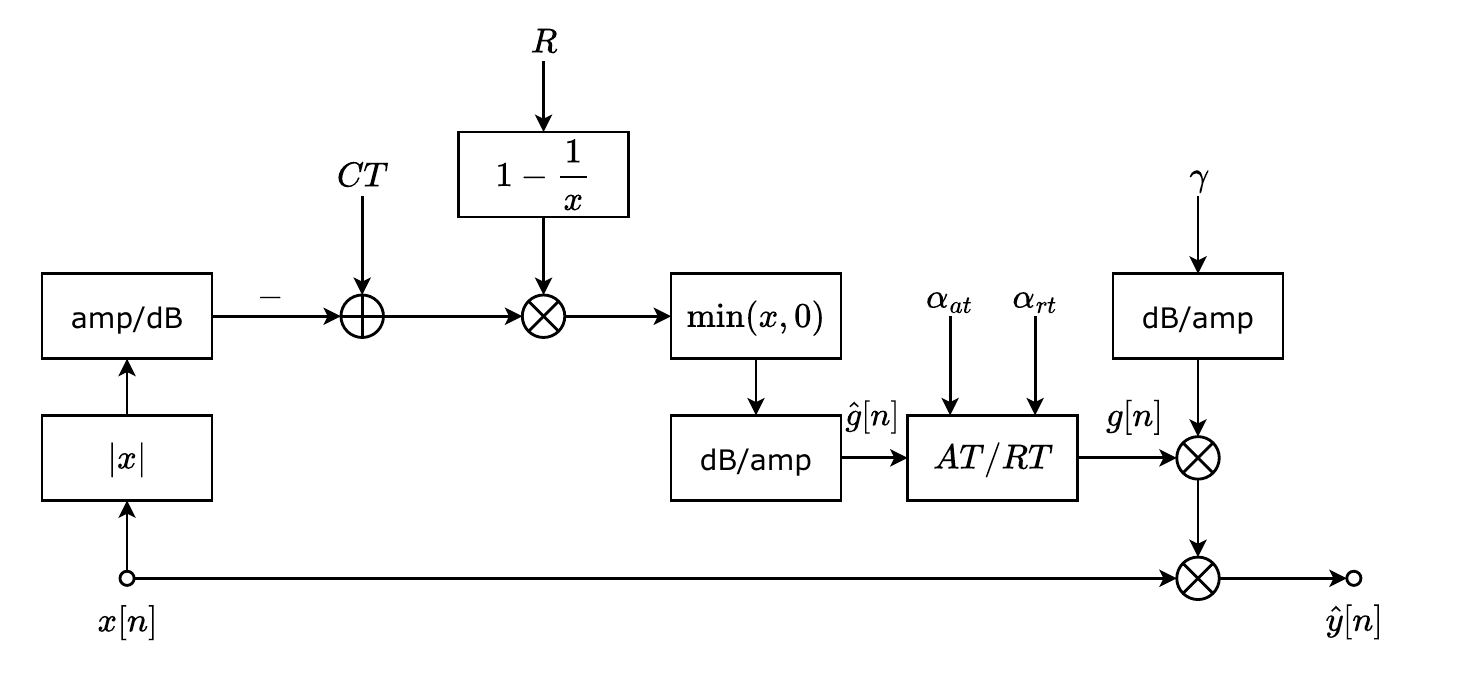}
    \caption{Block diagram of the feed-forward compressor. ``AT/RT'' refer to the ballistics control in Eq.~\eqref{eq:ballistics}.}
    \label{fig:ff_diagram}
\end{figure}

The resulting compressor consists of five parameters ($M = 5$): threshold ($CT$), ratio ($R$), attack time ($t_{at}$), release time ($t_{rt}$), and make-up gain ($\gamma$).
Threshold and make-up gain are expressed in decibels, and the attack/release times are in milliseconds.
Threshold controls the level at which the compressor starts to work, and the ratio determines the amount of gain reduction.
The make-up gain adjusts the output level to compensate for the gain reduction.
The attack and release time control the ballistics of the gain reduction, which is defined as:
\begin{multline}
    g[n] =  \\
    \begin{cases}
        \alpha_{at}\hat{g}[n] + (1 - \alpha_{at})g[n-1] & \text{if } \hat{g}[n] < g[n-1], \\
        \alpha_{rt}\hat{g}[n] + (1 - \alpha_{rt})g[n-1] & \text{otherwise},
    \end{cases}\label{eq:ballistics}
\end{multline}
\begin{equation}
    \alpha_{at/rt} = 1 - \exp\left(-\frac{2200T}{t_{at/rt}}\right).
\end{equation}
$\hat{g}[n]$ is the gain reduction before applying the ballistics.
Equation~\eqref{eq:ballistics} is a recursive smoothing filter, and the attack and release times determine how fast the filter responds to the loudness change.
The whole signal flow is shown in Fig.~\ref{fig:ff_diagram}.

Both threshold and make-up gain can be any real numbers, but not for the other parameters.
We also want to bound the parameters to some reasonable range that is common in practice.
To enforce these constraints, we define the parameter vector as $\theta = \{CT, \gamma, \hat{R}, \hat{\alpha}_{at}, \hat{\alpha}_{rt}\} \in \R^5$ and get
\begin{multline}
    \label{eq:parametrise}
    \{R, \alpha_{at}, \alpha_{rt}\} = \Big\{\lambda(u) \mid u \in  \{\hat{R}, \hat{\alpha}_{at}, \hat{\alpha}_{rt}\}, \Big.\\
    \Big.\lambda(u) = \theta_{min}(u) + \big(\theta_{max}(u) - \theta_{min}(u)\big)\sigma(u)\Big\},
\end{multline}
where $\sigma: \R \rightarrow [0, 1]$ is the monotonic sigmoid function, and $\theta_{min}: \R \rightarrow \R$ and $\theta_{max}: \R \rightarrow \R$ return the lower and upper bounds of the parameters, respectively.

\subsection{Newton-Raphson Optimisation}
\label{ssec:newton}
Newton-Raphson is a fundamental optimisation method that finds the minimum of a smooth and convex function $\mathcal{L}: \R^M \rightarrow \R$.
It updates the parameter $\theta$ by the following equation:
\begin{equation}
    \label{eq:newton}
    \theta \mapsto \theta - [\nabla^2 \mathcal{L}(\theta)]^{-1}\nabla \mathcal{L}(\theta).
\end{equation}
Here $\mathcal{L}(\theta) = \mathcal{D}(f_\x(\theta), \y)$, and $\nabla \mathcal{L}(\theta) \in \R^M$ and $\nabla^2 \mathcal{L}(\theta) \in \R^{M \times M}$ are the gradient and Hessian matrix of $\mathcal{L}$ at $\theta$, respectively.
NR leverages second-order information from the Hessian; thus, it guarantees convergence to a local minimum and converges faster than gradient descent, especially when the initial point is close to the minimum.

NR has two requirements that need to be satisfied:
\setlist{nolistsep}
\begin{enumerate}[noitemsep]
    
    \item $\mathcal{L}$ must be convex around the optimal $\theta^*$.
    \item $\mathcal{L}$ must be twice differentiable.
\end{enumerate}
We validate 1) empirically by observing the convergence behaviour, which we will discuss more in Section~\ref{ssec:optimisation}.
For 2), we set the distance function to be the square distance
$\mathcal{D}(f_\x(\theta), \y) = (f_\x(\theta) - \y)^\top(f_\x(\theta) - \y)$, which is twice differentiable.
Regarding the differentiability of the feed-forward compressor $f$, although \texttt{torchcomp} computes the sub-gradients for operations that are not differentiable everywhere (the absolute and minimum functions in Fig.~\ref{fig:ff_diagram} and the if-else statement in Eq.~\eqref{eq:ballistics}), it is shown that they are accurate enough for first-order optimisation.
Since its gradient backpropagation function is also differentiable, we can quickly evaluate its second-order derivatives within automatic differentiation frameworks like PyTorch~\citep{paszke_pytorch_2019}.

\section{Hessian Computation}
\label{sec:hessian}

\texttt{torchcomp} implements the Vector-Jacobian product (VJP)~\citep{rirsch1974differential,robeyns2023jvpvjp} $\vv \mapsto \vv^\top\nabla f_\x(\theta)$ of the compressor, which lets us compute the VJP of the loss function $\vv \mapsto \vv^\top\nabla \mathcal{L}(\theta)$ using backpropagation where $\nabla \mathcal{L}(\theta) = \nabla_{f_\x(\theta)} \mathcal{D}(f_\x(\theta), \y)\nabla_{\theta} f_\x(\theta)$.
We get the gradient by setting $\vv = 1$.

To get the Hessian matrix $\nabla^2 \mathcal{L}(\theta)$, note that the Hessian matrix is the Jacobian of $\nabla \mathcal{L}(\theta)$.
Given an identity matrix $\mathbf{I} \in \R^{M \times M}$, if we pass the $i^{\rm th}$ column of $\mathbf{I}$ into the VJP of $\nabla \mathcal{L}: \R^M \mapsto \R^M$, which is  $\vv \mapsto \vv^\top\nabla^2 \mathcal{L}(\theta)$, the result will be the $i^{\rm th}$ row of the Hessian $\nabla^2 \mathcal{L}(\theta)$; computing the entire Hessian equals passing every column of $\mathbf{I}$ to the VJP function in parallel ($\mathbf{I}^\top\nabla^2 \mathcal{L}(\theta)$).

Although we can get the Hessian using VJP twice (which is known as \emph{reverse-over-reverse} mode) as we show above, the second (sometimes also called outer) Jacobian can also be computed with Jacobian-Vector product (JVP) $\vv \mapsto \nabla h(\theta)\vv$~\citep{iclr2024howtocompute}.
This is called \emph{forward-over-reverse} mode where $h = \nabla \mathcal{L}$.
Since the Jacobian of $\nabla h(\theta)$ is a square matrix, computing it by passing the basis vectors, either from the right or the left, has to be repeated the same number of times.
The efficiency thus depends on the computational costs of VJP and JVP of $h: \R^M \rightarrow \R^M$, although in practice, the \emph{forward-over-reverse} strategy is usually preferred.
For a fair comparison, we derive the JVP of the function $\vv \mapsto \vv^\top\nabla f_\x(\theta)$, specifically, the backpropagation of the ballistics module in Eq.~\eqref{eq:ballistics}, as it involves recursive filters and is non-trivial to implement in PyTorch.

\subsection{Forward-Mode Gradients for Time-Varying One-Pole Filters}
\label{ssec:forward_all_pole}

To get the gradients of Eq.~\eqref{eq:ballistics}, let us rewrite it as a time-varying one-pole filter:
\begin{gather}
    \beta[n]      = 1 - \alpha_{at}^{\zeta[n]}\alpha_{rt}^{1 - \zeta[n]}, \label{eq:zeta_mask} \\
    \tilde{g}[n]  = (1 - \beta[n])\hat{g}[n],                               \label{eq:tilde_g} \\
    \textcolor{teal}{\underbrace{g[n]}_{\text{output}}} =
    \textcolor{blue}{\underbrace{\tilde{g}[n]}_{\text{input}}} +
    \textcolor{red}{\underbrace{\beta[n]}_{\text{multiplier}}} \mspace{10mu}
    \textcolor{teal}{\underbrace{g[n-1]}_{\text{previous output}}}, \label{eq:ballistics_rewritten}
\end{gather}
where $\zeta[n] \in \{0, 1\}$ is the attack and release phase indicator function.
We colour-coded Eq.~\eqref{eq:ballistics_rewritten} to emphasise the recursions as it will show up multiple times in the derivations, and we have special treatment for it in Section~\ref{ssec:acceleration}.

According to~\citep{ycy2024diffapf}, backpropagating $\nabla_{g[n]} \mathcal{L}(\theta)$ through Eq.~\eqref{eq:ballistics_rewritten} can be expressed as passing through a time-varying one-pole filter in reversed time:
\begin{equation}
    \textcolor{teal}{\nabla_{\tilde{g}[n]} \mathcal{L}(\theta)} = \textcolor{blue}{\nabla_{g[n]} \mathcal{L}(\theta)} + \textcolor{red}{\beta[n+1]}\textcolor{teal}{\nabla_{\tilde{g}[n+1]} \mathcal{L}(\theta)}.
    \label{eq:tilde_g_grad}
\end{equation}
The gradients $\nabla_{\beta[n]} \mathcal{L}(\theta)$ can be computed using chain rule:
\begin{equation}
    \nabla_{\beta[n]} \mathcal{L}(\theta) = \nabla_{\tilde{g}[n]} \mathcal{L}(\theta)(g[n-1] - \hat{g}[n]). \label{eq:beta_grad}
\end{equation}
The gradients $\nabla_{\alpha_{at}} \mathcal{L}(\theta)$, $\nabla_{\alpha_{rt}} \mathcal{L}(\theta)$, and $\nabla_{\hat{g}[n]} \mathcal{L}(\theta)$ are trivial to compute given Eq.~\eqref{eq:zeta_mask} and Eq.~\eqref{eq:tilde_g}.

Backpropagation of $\nabla_{\tilde{g}[n]} \nabla \mathcal{L}(\theta)$ (we simplify $\nabla_{\nabla_{*}\mathcal{L}(\theta)}$ as $\nabla_{*}$ for brevity) through Eq.~\eqref{eq:tilde_g_grad} can be done in the same way as in Eq.~\eqref{eq:ballistics_rewritten}:
\begin{gather}
    \textcolor{teal}{\nabla_{g[n]} \nabla \mathcal{L}(\theta)} = \textcolor{blue}{\nabla_{\tilde{g}[n]} \nabla \mathcal{L}(\theta)} + \textcolor{red}{\beta[n]}\textcolor{teal}{\nabla_{g[n-1]} \nabla \mathcal{L}(\theta)},\label{eq:tilde_g_hessian} \\
    \nabla_{\beta[n+1]} \nabla \mathcal{L}(\theta) = \nabla_{g[n]} \nabla \mathcal{L}(\theta)\nabla_{\tilde{g}[n+1]} \mathcal{L}(\theta).    \label{eq:beta_hessian}
\end{gather}
For forward propagation (JVP) on Eq.~\eqref{eq:tilde_g_grad}, let us use Eq.~\eqref{eq:ballistics_rewritten} as an example.
Since in the forward mode the Jacobians are multiplied from right to left (opposite to the backpropagation), we need to apply Eq.~\eqref{eq:tilde_g_grad} and Eq.~\eqref{eq:beta_grad} in reverse order and traverse the computational graph from the outputs back to the inputs.
This also means applying recursive filters in the opposite direction.
Given forwarded gradients $\nabla_{\theta} \tilde{g}[n]$ and $\nabla_{\theta} \beta[n]$, the forward version of Eq.~\eqref{eq:tilde_g_grad} is:
\begin{multline}
    \textcolor{teal}{\nabla_{\theta} g[n]} = \textcolor{blue}{\left[\nabla_{\theta} \tilde{g}[n] + \nabla_{\theta} \beta[n] \left(g[n-1] - \hat{g}[n]\right)\right]} \\
    + \textcolor{red}{\beta[n]}\textcolor{teal}{\nabla_{\theta} g[n-1]},
    \label{eq:tilde_g_grad_forward}
\end{multline}
which is just changing the input of Eq.~\eqref{eq:ballistics_rewritten} to $\nabla_{\theta} \tilde{g}[n] + \nabla_{\theta} \beta[n] (g[n-1] - \hat{g}[n])$.
Applying the same process to Eq.~\eqref{eq:tilde_g_grad}, we can get the forward version of Eq.~\eqref{eq:tilde_g_hessian} similar to Eq.~\eqref{eq:tilde_g_grad_forward} as:
\begin{multline}
    \textcolor{teal}{\nabla_{\theta} \nabla_{\tilde{g}[n]} \mathcal{L}(\theta)} = \\
    \textcolor{blue}{[\nabla_{\theta} \nabla_{g[n]} \mathcal{L}(\theta) + \nabla_{\theta} \beta[n] \nabla_{\tilde{g}[n+1]} \mathcal{L}(\theta)]} \\
    + \textcolor{red}{\beta[n+1]}\textcolor{teal}{\nabla_{\theta} \nabla_{\tilde{g}[n+1]} \mathcal{L}(\theta)}.
    \label{eq:tilde_g_hessian_forward}
\end{multline}

Comparing the VJP of Eq.~\eqref{eq:tilde_g_grad} (Eq.~\eqref{eq:tilde_g_hessian} and Eq.~\eqref{eq:beta_hessian} combined) to its JVP (Eq.~\eqref{eq:tilde_g_hessian_forward}), we see that the JVP has one more addition than the VJP.
Nevertheless, the extra addition does not occur within the recursion, and modern computers are very good at vectorised atomic addition, so the computation time is often negligible.

In summary, in this section, we show that propagating the gradients of a time-varying one-pole filter in forward mode equals filtering the gradients with the same filter\footnote{This result generalises to higher order all-pole filters as well.}.
The resulting Eq.~\eqref{eq:tilde_g_grad_forward} is required for the JVP of the inner Jacobian $\vv \mapsto \nabla \mathcal{L}(\theta)\vv$, while Eq.~\eqref{eq:tilde_g_hessian_forward} is required for the JVP of the outer Jacobian $\vv \mapsto \nabla h(\theta)\vv$.

\section{Experiment}
\label{sec:exp}

\subsection{Dataset}
\label{ssec:dataset}
We use the recordings from SignalTrain~\citep{colburn_2020_3824876} as it is the largest curated analogue compressor dataset known to the authors and has been utilised in many previous works~\citep{steinmetz_efficient_2022, wright2022grey, ycy2024diffapf, yeh_hyper_2024}.
It contains paired recordings sampled at \SI{44.1}{\kHz} from LA-2A.
Technically, the LA-2A is a levelling amplifier (LA) and does not have the parameters mentioned in Section~\ref{ssec:ff}.
It has only one knob, called \textit{peak reduction}, that controls the compression.
The peak reduction was sampled from 0 to 100 with a spacing of 5, and the gain value was fixed for the compressor and limiter modes.
We select recordings labelled with \texttt{3c} in the file name, which use an identical \SI{20}{\minute} audio stimulus consisting of both musical and synthetic sounds for all configurations.

\subsection{Acceleration Techniques}
\label{ssec:acceleration}
We implement everything in PyTorch.
To accelerate the training process, we split the audio into \SI{12}{\second} chunks with \SI{1}{\second} overlap, so $f_\x$, which is inherently sequential in contrast to $\nabla f_\x$, can be parallelised at the batch level.
The overlap parts are used as a warm-up, and the loss is computed only on each chunk's last \SI{11}{\second}.
Following~\citep{wright2022grey}, we apply a pre-emphasis filter $\frac{1 - z^{-1}}{1 - 0.995 z^{-1}}$ to $\y$ and $\haty$ prior to computing the loss to reduce low-frequency noise.

Furthermore, we identify that a lot of the operations are one-pole filters (Eq.~\eqref{eq:ballistics_rewritten}), and a parallel algorithm exists for these filters called \textit{parallel associative scan}~\citep{BlellochTR90}.
The idea is to convert the recursion into a form like $s[0] \oplus s[1] \oplus \ldots \oplus s[n]$ where $\oplus$ is associative, then the scan algorithm can be applied.
We use the CUDA implementation from~\citep{martin2018parallelizing} to parallelise the coloured equations (except Eq.~\eqref{eq:ballistics_rewritten}) and the pre-emphasis filter on the GPU.

\subsection{Analysis on Computational Cost}
PyTorch has two implementations for computing the Hessian.
One lies under the \texttt{autograd} module, which is the default implementation and has been around since PyTorch 1.5.
It only has full coverage for \emph{reverse-over-reverse} (hereafter referred to as \emph{rev-rev}) mode.
The other implementation is the \texttt{func} module, introduced in PyTorch 2.0, which is relatively new but allows for the flexible composition of forward and reverse mode Jacobians.
We evaluate all the four combination strategies for the Hessian $\nabla^2 \mathcal{L}(\theta)$ using \texttt{func} and compare them to the default \texttt{autograd} implementation.

\begin{table}[h]
    \caption{The memory cost (in MB) and runtime (in milliseconds) of different hessian computation strategies in PyTorch.}
    \centering
    \resizebox{\columnwidth}{!}{
        \begin{tabular}{cc|cccc}
            \toprule
                   & \texttt{autograd} & \multicolumn{4}{c}{\texttt{func}}                               \\
                   & rev-rev           & rev-rev                           & fwd-rev & rev-fwd & fwd-fwd \\
            \midrule
            Memory & $\mathbf{1066}$   & 2358                              & 3534    & 6306    & 5364    \\
            Time   & $\mathbf{26.5}$   & 28.1                              & 28.1    & 64.7    & 38.9    \\
            \bottomrule
        \end{tabular}
    }
    \label{tab:torchfunc_analysis}
\end{table}

We report the computational cost in Table~\ref{tab:torchfunc_analysis}, measured on an RTX 3060 with a batch size of 16.
From Table~\ref{tab:torchfunc_analysis}, we see that the \emph{rev-rev} and \emph{fwd-rev} modes are equally fast, which implies that the extra addition we mentioned in Section~\ref{sec:hessian} does not impact the performance.
The \emph{rev-fwd} mode is the slowest and most memory-consuming.
We see that \emph{fwd-rev} requires more memory than \emph{rev-rev}, probably due to non-optimal memory utilisation for the forward gradients in our implementation\footnote{In our filter implementation, only one dimension of the tensor is treated as batch dimension. The straightforward way to call the JVP M times in parallel using the same function is to duplicate the parameters M times along the batch dimension.}.
The table also shows inherent overhead in memory and time for the \texttt{func} module, as the numbers are all bigger than those of \texttt{autograd}.
This is likely because this feature is still in the Beta stage and has not been optimised enough.

Based on the above findings, we choose the \texttt{autograd} implementation.
All the computations for each configuration can be fitted on a single RTX 3060 with 12GB of VRAM, and no mini-batching is required.
The resulting optimisation takes roughly just \SI{3.5}{\second} for one NR update.
Moreover, training on the entire dataset takes less than \SI{20}{\minute}, instead of hours as reported in~\citep{wright2022grey}.

\subsection{Optimisation Strategy}
\label{ssec:optimisation}
We perform the damped NR method with backtracking line search~\citep{more1982newton}.
In each step, the parameters are updated as $\theta \mapsto \theta - \tau \mathbf{\nu}$, where $\tau \in [0, 1]$ is the step size and $\mathbf{\nu}$ is the solution of $\nabla^2 \mathcal{L}(\theta)\mathbf{\nu} = \nabla \mathcal{L}(\theta)$.
We use an equation solver to avoid performing matrix inversion for better numerical stability.
Starting from $\tau = 1$, we divide $\tau$ by 2 until the Armijo–Goldstein condition~\citep{Armijo1966} $\mathcal{L}(\theta - \tau \mathbf{\nu}) \leq \mathcal{L}(\theta) - \alpha \tau \nabla \mathcal{L}(\theta)^\top \mathbf{\nu}$ is satisfied with $\alpha = 0.0001$.
If the Hessian matrix is not positive semi-definite (which implies negative curvature), we randomly sample a new direction orthogonal to $\mathbf{\nu}$ and repeat the process.

We find different optimal parameters for each sampled peak reduction.
Since the NR method is sensitive to the initialisation, we start from 100 peak reduction which has the heaviest compression so it is easier to fit, and then use the previous result to initialise the subsequent peak reduction, e.g., 100 $\rightarrow$ 95 $\rightarrow$ 90, etc.
This ensures the starting point for each configuration is as close to the solution as possible.
The initial values are determined empirically based on early experiments and shared beliefs of the target circuit, which are $CT = -36$ dB, $\gamma = 0$ dB, $R = 4$, $t_{at} = 1$ ms, and $t_{rt} = 200$ ms.
We restrict the range of the parameters to $R \in [1, 20]$, $t_{at} \in [0.1, 100]$ ms, and $t_{rt} \in [10, 1000]$ ms.

The optimisation converges in less than 10 iterations in most cases.
We stopped the process at 40 peak reduction, as the NR method could not find a consistent solution for lower peak reductions.
The optimisation encounters negative curvature structure frequently~\footnote{For peak reductions over 40, this rarely happens (it happens a few times when random initialisation is used).}.
We investigated the data and found that the compression is less noticeable in this range and is surpassed by other non-linearities that a feed-forward compressor cannot capture.
For other peak reductions, the NR method converges to the same solution in multiple different initialisations, verifying that $\mathcal{L}(\theta)$ is mostly convex around the optimal solution.

\subsection{Results}
\label{ssec:results}

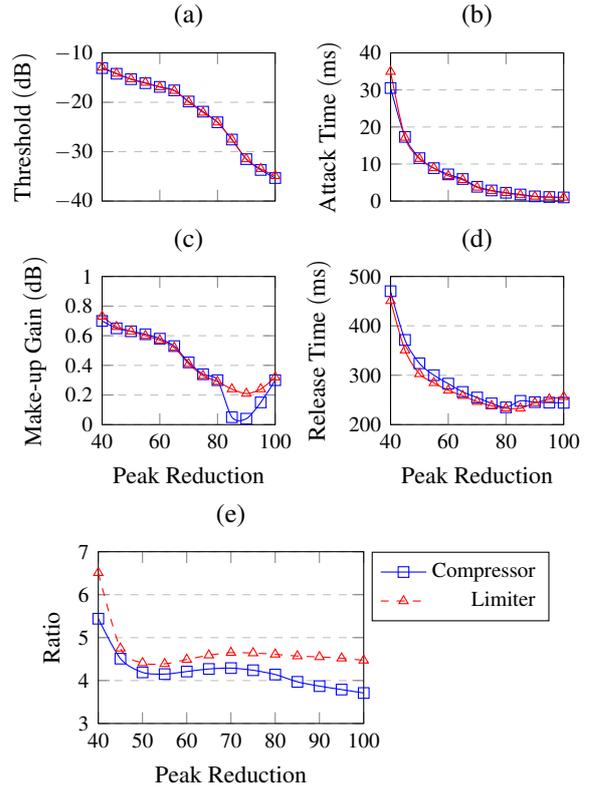
\begin{figure}[h]
    \centering
    \subimport{figures/}{peak2param.tex}
    \caption{The mapping from LA-2A peak reduction to the compressor's parameters.
    }
    \label{fig:mapping}
\end{figure}

\begin{quoting}
    \textit{...The average compression ratio is always set at roughly 3:1, while the average Attack time is 10 milliseconds, and the Release time is about 60 milliseconds for 50\% of the release, and anywhere from 1 to 15 seconds for the rest.}~\footnote{\href{https://www.uaudio.com/blog/la-2a-collection-tips-tricks/}{www.uaudio.com/blog/la-2a-collection-tips-tricks}}
\end{quoting}
The learnt configurations are shown in Fig.~\ref{fig:mapping}.
The mappings to the parameters are nearly identical between the compressor and limiter modes, except for the ratio, which is slightly higher in the limiter mode but exhibits a similar trend.
The ratio is typically around 4:1 in compressor mode, slightly deviating from the manufacturer's stated ratio of 3:1.
The attack and release times vary exponentially with the peak reduction, rather than fixed values.
Since there is only one release time parameter, it does not capture the documented two-stage release behaviour of the LA-2A, but an average of them.

\section{Evaluation}
\label{sec:eval}

We compare our compressor (denoted as 4A-2A and ``4'' stands for feed-forward) with the following baselines: the official UAD LA-2A plugin~\footnote{\href{https://www.uaudio.com/uad-plugins/compressors-limiters/teletronix-la-2a-tube-compressor.html}{www.uaudio.com/uad-plugins/compressors-limiters/teletronix-la-2a-tube-compressor}}, the emulation plugin CA-2A by Cakewalk~\footnote{\href{https://legacy.cakewalk.com/Products/CA-2A}{legacy.cakewalk.com/Products/CA-2A}}, the emulation plugin CLA-2A by Waves~\footnote{\href{https://www.waves.com/plugins/cla-2a-compressor-limiter}{www.waves.com/plugins/cla-2a-compressor-limiter}}, and the grey-box compressor from~\citep{wright2022grey}.
We evaluate 4A-2A using its training data, as the five-parameter model cannot fully represent the LA-2A and no overfitting is observed on this dataset with more complex models~\citep{wright2022grey}.
We evaluate performance using the Error-to-Signal Ratio (ESR) and the Loudness Dynamic Range difference ($\Delta$LDR).
ESR is defined as $\frac{(\y - \hat{\y})^\top(\y - \hat{\y})}{\y^\top \y}$.
$\Delta$LDR is the difference between the LDR~\citep{nercessian_direct_2022} of the target signal and the predicted signal.
The same pre-emphasis filter is applied to all audio before evaluations.

To calculate LDR given $\y$, we first compute the log ratio between the short-term and long-term RMS envelopes as
\begin{equation}
    L_\y[n] = 10 \log_{10}\frac{\text{RMS}_\text{short}(y[n])}{\text{RMS}_\text{long}(y[n])}.
\end{equation}
The integration time for the short-term RMS is set to \SI{50}{\milli\second} and the long-term RMS is set to \SI{3}{\second} where the details can be found in~\citep{nercessian_direct_2022}.
We define $\text{LDR}(\y)$ as
\begin{equation}
    \text{LDR}(\y) = \sqrt{\frac{1}{N}\sum_{n=0}^{N-1}L_\y[n]^2},
\end{equation}
which tells how much the loudness varies locally (microdynamics) in $\y$.
$\Delta\text{LDR}$ is simply $\text{LDR}(\hat{\y}) - \text{LDR}(\y)$ so if $\Delta\text{LDR} > 0$, it means $\hat{\y}$ is less compressed than $\y$ and vice versa.

Since we do not know what setup and gain staging were used to model the commercial plugins, the plugins' input and output gains must be tuned.
We manually adjust the input gain so that the sum of its $\Delta$LDR across peak reductions is close to zero, ensuring its compression curve matches the SignalTrain dataset.
We then adjust the output gain so that its overall ESR is minimised.
We render the audio through plugins using \texttt{pedalboard}~\citep{sobot_peter_2023_7817838}.

We choose the Model 5 configuration for the grey-box compressor, their best-performing model.
It is architecture-wise very similar to our 4A-2A, but the make-up gain $\gamma$ is replaced with a GRU.
Thus, we follow their setup to train a GRU on the same training set but pre-processed by 4A-2A with $\gamma = $ \SI{0}{dB}.
We set the batch size to 26 (13 configurations $\times$ two modes) and the sequence length to 12288.
The training converges after around eight epochs.
We stop the training at the $15^{\rm th}$ epoch and pick the checkpoint with the lowest loss.
The process takes less than 30 minutes on an RTX 3060 Laptop GPU.
We denote this model as 4A-2A-G.

\begin{figure}[h]
    \centering
    \subimport{figures/}{esr.tex}
    \caption{ESR and $\Delta$LDR of all the evaluated models under different peak reduction and mode.}
    \label{fig:metrics}
\end{figure}
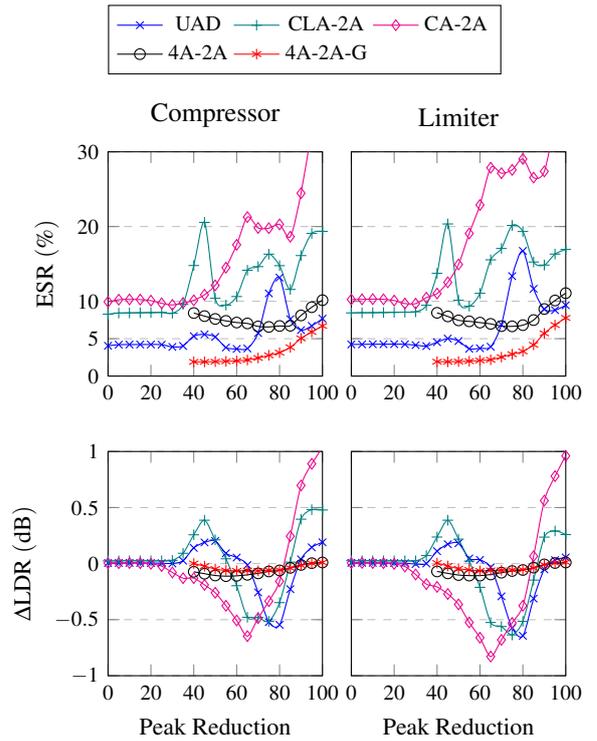

The evaluation results are shown in Fig.~\ref{fig:metrics}.
4A-2A-G has the lowest ESR and is close to those reported in~\citep{wright2022grey}.
4A-2A shows comparable performance to the official UAD plugin.
At around 75 peak reduction, the LA-2A behaves most like the feed-forward compressor, thus having the lowest ESR.
Beyond this point, the 4A-2A's ESR increases, and the errors are more for the limiter mode.
For the commercial plugins, their curves are less smooth and have a few apparent spikes.
Among them, the CA-2A has the highest ESR in general and even reaches over \SI{40}{\percent} at 100 peak reduction.

Regarding their $\Delta$LDR profiles, we see that CA-2A varies the most with a maximum mismatch of \SI{1}{\dB} at 100 peak reduction.
UAD and CLA-2A exhibit similar patterns of difference, which can be divided into three stages: from 40 to 60 (peak reduction), their compressions are lighter than the ground truth; then, the compressions become heavier from 60 to 80 and then return to lighter levels beyond 80.
4A-2A and 4A-2A-G have their $\Delta$LDR very close to zero thus out-performing others, where 4A-2A-G being the closest.

\section{Application: the 4A-2A Plugin}
\label{sec:plugin}

\begin{figure}[h]
    \centering
    \includegraphics[width=0.9\columnwidth]{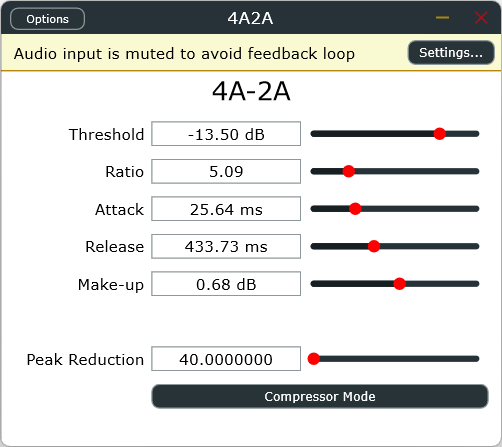}
    \caption{The GUI of the 4A-2A plugin.}
    \label{fig:gui}
\end{figure}

We implement our 4A-2A compressor as a real-time audio plugin using JUCE~\citep{juce}, providing it as a creative tool for audio engineers.
In the plugin interface (Fig.~\ref{fig:gui}), we have five sliders that control the parameters of the underlying compressor.
At the bottom, there is one slider for peak reduction and a switch for the mode.
The peak reduction slider is linked to the above compressor's parameters, mapping the peak reduction to them using the learnt mappings from Fig.~\ref{fig:mapping}.
The mode switch toggles between the compressor and limiter profiles.
The user can adjust the peak reduction and observe the compressor's behaviour in real-time.
If they want to fine-tune the compressor, they can adjust the parameters manually.

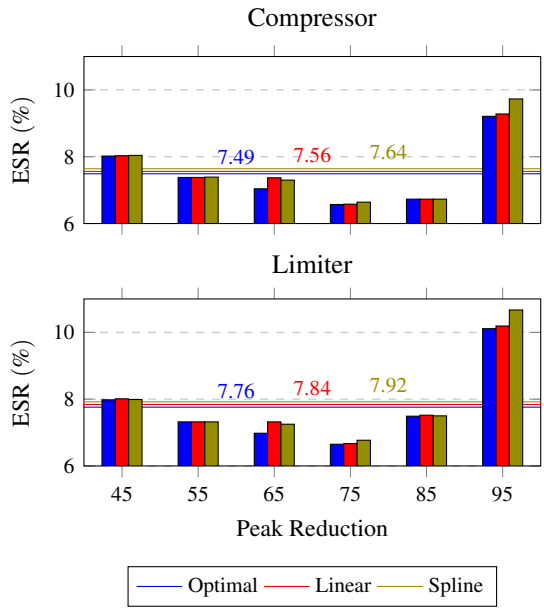
\begin{figure}
    \centering
    \subimport{figures/}{interp.tex}
    \caption{Comparison of different interpolation methods. The horizontal lines are the averages.}
    \label{fig:interp}
\end{figure}

The peak reductions in SignalTrain are sampled on a grid.
Interpolation is required for continuous mapping $[40, 100] \rightarrow \R^5$.
To evaluate the interpolation error, we exclude peak reductions $\{45, 55, \dots, 95\}$ and use the remaining 7 points to interpolate them using either linear or cubic spline interpolation.
Their ESR is shown in Fig.~\ref{fig:interp}.
Linear interpolation and spline interpolation perform the worst at 65 and 95 peak reduction, respectively.
On average, the spline interpolation has a slightly higher ESR than the linear interpolation.
We suppose this finding generalises to denser grids and choose the linear interpolation for the plugin.

We made the GRU model from 4A-2A-G into a standalone plugin.
The GRU model is very lightweight, consisting of only one layer with eight hidden units.
We wrapped the GRU model as a PyTorch script module using Neutone~\citep{fyfe2022neutone}, which can be loaded into their Neutone FX plugin~\footnote{\href{https://github.com/Neutone/neutone_sdk}{github.com/Neutone/neutone\_sdk}}.
Cascading the 4A-2A and the GRU model resembles the 4A-2A-G baseline, a real-time implementation of the Model 5 emulator from~\citep{wright2022grey}, with the feed-forward part being trained using the NR method, outperforming all commercial plugins in our evaluation.

\begin{figure*}[!t]
    \centering
    \includegraphics[width=\textwidth]{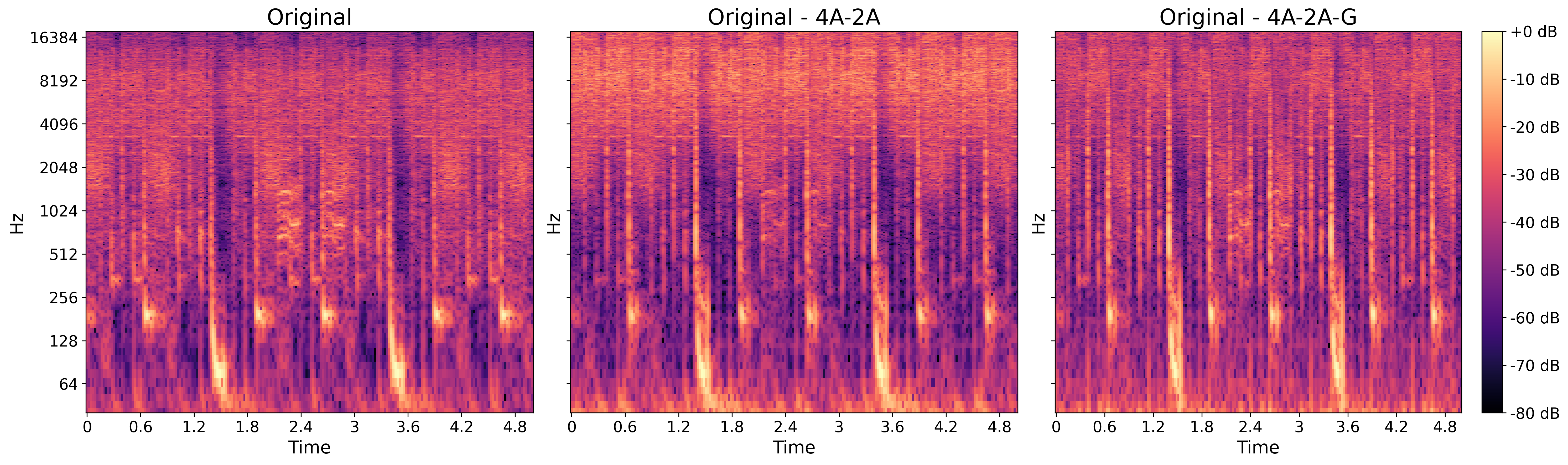}
    \caption{The spectrograms of a drum loop from the SignalTrain and the residuals ($\y - \hat{\y}$) of the model's predictions with 75 peak reductions operated in compressor mode.}
    \label{fig:spec}
\end{figure*}

Comparing the residuals with or without the GRU make-up gain shown in Fig.~\ref{fig:spec}, the most noticeable difference is in the high-frequency range (> \SI{4}{\kHz}).
The GRU successfully captures the non-linearities of these analogue circuits, which cannot be modelled by a simple digital compressor, and spreads out the errors across frequencies.

\section{Discussion}
\label{sec:discussion}

The NR method effectively finds a unique solution to match the LA-2A's response to the square distance criterion.
There is room for improvement, as LA-2A is technically a feedback compressor; therefore, the signal flows of 4A-2A and LA-2A are not precisely matched.
By using a digital feedback compressor ($s[n] = \hat{y}[n-1]$) and modelling the two-stage release behaviour explicitly, we can potentially improve the performance.
However, efficiently backpropagating gradients through the feedback path is challenging and requires custom implementations.
In contrast, our grey-box sound-matching approach can be generalised to any dynamic range controller circuit, and our trial on the LA-2A already yields satisfactory results.

Due to the variations between analogue circuits, the inferior plugin performance we see in Section~\ref{sec:eval} is expected.
Their modelled LA-2A might behave slightly differently from the SignalTrain one.
It is interesting to note that the UAD and CLA-2A have similar ESR/$\nabla$LDR patterns, which could be due to similar target circuits or modelling approaches.

We evaluate the entire Hessian matrix, as the number of parameters is smaller.
However, for more complex effects and longer effect chains, the Hessian can become too large to fit in memory.
Methods like conjugate gradients using Hessian-vector product $\vv \mapsto \nabla^2 \mathcal{L}(\theta)\vv$ can be used to solve the equation $\nabla^2 \mathcal{L}(\theta)\mathbf{\nu} = \nabla \mathcal{L}(\theta)$ without explicitly computing the Hessian~\citep{iclr2024howtocompute, hestenes1952methods}.
Nevertheless, increasing the number of parameters also raises the possibility of encountering more local minima, making it harder to find the optimal solution.

\section{Conclusion}
\label{sec:conclusion}

This paper presented a method to emulate the LA-2A optical compressor using a feed-forward digital compressor with parameters optimised via the Newton-Raphson method.
Our approach successfully turned the LA-2A's response into interpretable parameters that can be used for creative control.
The efficient implementation of ballistics and the parallel associative scan algorithm for gradient computation significantly accelerate optimisation.
Forward-mode gradient propagation through the compressor is derived to benchmark different strategies for computing the Hessian.

Our findings suggest that the Newton-Raphson method is a promising approach for the grey-box modelling of audio compressors, providing a foundation for further research and development in this area.
Further improvements could be made by using a feedback design and explicitly modelling the LA-2A's two-stage release behaviour.
Future work could extend Newton-Raphson optimisation to other analogue compressors or non-linear audio effects.
Another direction could be reducing the computational costs of the Hessian matrix for more complex signal chains.

\section*{Acknowledgements}
The first author is a research student supported jointly by UKRI (grant number EP/S022694/1) and Queen Mary University of London.

\bibliographystyle{jaes}

\bibliography{refs}

\end{document}

%% file: figures/peak2param.tex
\begin{tikzpicture}[scale=1]
    \begin{groupplot}[group style={
                    group size=2 by 2,
                    horizontal sep=0.2\columnwidth,
                    x descriptions at=edge bottom,
                },
            width=0.3\columnwidth,
            xlabel={\small Peak Reduction},
            xmin=40, xmax=100,
            tick label style={font=\footnotesize},
            ymajorgrids=true,
            grid style=dashed,
            scale only axis,
        ]
        \nextgroupplot[
            ylabel={\small Threshold},
            y unit= dB,
            ymin=-40, ymax=-10,
            title={(a)}
        ]
        \addplot[
            smooth,
            color=blue,
            mark=square,
        ]
        coordinates {
                (40, -13.08)(45, -14.22)(50, -15.32)(55, -16.13)(60, -16.88)(65, -17.59)(70, -19.84)(75, -21.90)(80, -24.05)(85, -27.55)(90, -31.51)(95, -33.70)(100, -35.34)
            };
        \addplot[
            smooth,
            color=red,
            mark=triangle,
            mark options={solid},
        ]
        coordinates {
                (40, -12.93)(45, -14.17)(50, -15.30)(55, -16.01)(60, -16.92)(65, -17.65)(70, -19.92)(75, -22.06)(80, -24.04)(85, -27.53)(90, -31.49)(95, -33.48)(100, -34.90)
            };

        \nextgroupplot[
            ylabel={\small Attack Time},
            y unit= ms,
            ymin=0, ymax=40,
            title={(b)}
        ]
        \addplot[
            smooth,
            color=blue,
            mark=square,
        ]
        coordinates {
                (40, 30.52)(45, 17.32)(50, 11.58)(55, 8.88)(60, 7.20)(65, 5.95)(70, 3.85)(75, 2.86)(80, 2.24)(85, 1.77)(90, 1.27)(95, 1.09)(100, 0.98)
            };
        \addplot[
            smooth,
            color=red,
            mark=triangle,
            mark options={solid},
        ]
        coordinates {
                (40, 34.95)(45, 17.14)(50, 11.43)(55, 8.92)(60, 6.80)(65, 5.72)(70, 3.68)(75, 2.74)(80, 2.21)(85, 1.64)(90, 1.23)(95, 1.06)(100, 0.97)
            };
        \nextgroupplot[
            ylabel={\small Make-up Gain},
            y unit= dB,
            ymin=0, ymax=1,
            title={(c)}
        ]
        \addplot[
            smooth,
            color=blue,
            mark=square,
        ]
        coordinates {
                (40, 0.70)(45, 0.65)(50, 0.63)(55, 0.61)(60, 0.58)(65, 0.53)(70, 0.42)(75, 0.34)(80, 0.30)(85, 0.05)(90, 0.04)(95, 0.15)(100, 0.30)
            };
        \addplot[
            smooth,
            color=red,
            mark=triangle,
            mark options={solid},
        ]
        coordinates {
                (40, 0.73)(45, 0.66)(50, 0.63)(55, 0.60)(60, 0.57)(65, 0.52)(70, 0.41)(75, 0.33)(80, 0.29)(85, 0.24)(90, 0.21)(95, 0.24)(100, 0.32)
            };
        \nextgroupplot[
            ylabel={\small Release Time},
            y unit= ms,
            ymin=200, ymax=500,
            title={(d)}
        ]
        \addplot[
            smooth,
            color=blue,
            mark=square,
        ]
        coordinates {
                (40, 470.18)(45, 371.46)(50, 324.13)(55, 300.07)(60, 283.02)(65, 266.18)(70, 254.91)(75, 242.99)(80, 234.94)(85, 248.11)(90, 245.49)(95, 244.56)(100, 244.13)
            };
        \addplot[
            smooth,
            color=red,
            mark=triangle,
            mark options={solid},
        ]
        coordinates {
                (40, 450.68)(45, 350.75)(50, 302.67)(55, 284.56)(60, 269.35)(65, 258.54)(70, 246.36)(75, 238.02)(80, 233.05)(85, 233.05)(90, 243.92)(95, 251.62)(100, 256.32)
            };
    \end{groupplot}
\end{tikzpicture}
\begin{tikzpicture}[scale=1]
    \begin{axis}[
            name=ratio_plot,
            title={(e)},
            anchor=above north west,
            xlabel={\small Peak Reduction},
            ylabel={\small Ratio},
            xmin=40, xmax=100,
            ymin=3, ymax=7,
            xtick={40, 50, 60, 70, 80, 90, 100},
            ymajorgrids=true,
            grid style=dashed,
            scale only axis,
            width=0.46\columnwidth,
            height=0.3\columnwidth,
            tick label style={font=\footnotesize},
            legend style={
                    cells={anchor=east},
                    legend pos=outer north east,
                    font=\footnotesize,
                },
        ]
        \addplot[
            smooth,
            color=blue,
            mark=square,
        ]
        coordinates {
                (40, 5.44)(45, 4.51)(50, 4.19)(55, 4.15)(60, 4.21)(65, 4.27)(70, 4.29)(75, 4.24)(80, 4.14)(85, 3.97)(90, 3.87)(95, 3.79)(100, 3.71)
            };
        \addplot[
            smooth,
            color=red,
            mark=triangle,
            mark options={solid},
            dashed
        ]
        coordinates {
                (40, 6.51)(45, 4.75)(50, 4.41)(55, 4.39)(60, 4.49)(65, 4.59)(70, 4.65)(75, 4.64)(80, 4.61)(85, 4.57)(90, 4.55)(95, 4.52)(100, 4.47)
            };
        \legend{Compressor, Limiter}
    \end{axis}
\end{tikzpicture}

%% file: figures/esr.tex
\begin{tikzpicture}[scale=1]
    \begin{groupplot}[group style={
                    group size=2 by 2,
                    horizontal sep=0.05\columnwidth,
                    xlabels at=edge bottom,
                    y descriptions at=edge left,
                },
            width=0.58\columnwidth,
            height=0.6\columnwidth,
            xlabel={\small Peak Reduction},
            tick label style={font=\footnotesize},
            ymajorgrids=true,
            grid style=dashed,
        ]
        \nextgroupplot[
            title={\normalsize Compressor},
            ymin=0, ymax=30, ytick={0, 5, 10, 20, 30}, xmin=0, xmax=100, xtick={0, 20, 40, 60, 80, 100},
            ylabel={\small ESR},
            y unit= \%,
            legend style={legend columns=3, font=\footnotesize,
                    at={(0, 1.5)}, anchor=west},
        ]
        \addplot[smooth, color=blue, mark=x, mark size=2pt] coordinates {
                (0, 4.05) (5, 4.20) (10, 4.21) (15, 4.22) (20, 4.22) (25, 4.20) (30, 3.90) (35, 4.03) (40, 5.34) (45, 5.57) (50, 5.23) (55, 3.82) (60, 3.63) (65, 3.69) (70, 5.75) (75, 11.06) (80, 13.17) (85, 7.61) (90, 6.15) (95, 6.71) (100, 7.68)
            };
        \addlegendentry{UAD}
        \addplot[smooth, color=teal, mark=+, mark size=2pt] coordinates {
                (0, 8.28) (5, 8.43) (10, 8.45) (15, 8.47) (20, 8.49) (25, 8.51) (30, 8.41) (35, 9.60) (40, 14.79) (45, 20.56) (50, 10.40) (55, 9.49) (60, 10.64) (65, 14.16) (70, 14.64) (75, 16.31) (80, 14.77) (85, 11.59) (90, 16.12) (95, 19.08) (100, 19.36)
            };
        \addlegendentry{CLA-2A}
        \addplot[smooth, color=magenta, mark=diamond, mark size=2pt] coordinates {
                (0, 9.93) (5, 10.22) (10, 10.25) (15, 10.23) (20, 10.09) (25, 9.73) (30, 9.52) (35, 9.79) (40, 10.13) (45, 10.88) (50, 12.13) (55, 14.50) (60, 17.56) (65, 21.27) (70, 19.81) (75, 19.78) (80, 20.29) (85, 18.64) (90, 24.46) (95, 32.92) (100, 41.13)
            };
        \addlegendentry{CA-2A}
        \addplot[smooth, color=black, mark=o, mark size=2pt] coordinates {
                (40, 8.43) (45, 8.02) (50, 7.64) (55, 7.38) (60, 7.16) (65, 7.04) (70, 6.64) (75, 6.57) (80, 6.70) (85, 6.73) (90, 8.11) (95, 9.21) (100, 10.15)
            };
        \addlegendentry{4A-2A}
        \addplot[smooth, color=red, mark=asterisk, mark size=2pt] coordinates {
                (40, 1.90) (45, 1.90) (50, 1.93) (55, 1.97) (60, 2.03) (65, 2.15) (70, 2.42) (75, 2.77) (80, 3.13) (85, 3.84) (90, 5.04) (95, 5.90) (100, 6.70)
            };
        \addlegendentry{4A-2A-G}

        \nextgroupplot[title={\normalsize Limiter}, ymin=0, ymax=30, ytick={0, 5, 10, 20, 30}, xmin=0, xmax=100, xtick={0, 20, 40, 60, 80, 100}]
        \addplot[smooth, color=blue, mark=x, mark size=2pt] coordinates {
                (0, 4.23) (5, 4.24) (10, 4.26) (15, 4.26) (20, 4.27) (25, 4.25) (30, 4.14) (35, 4.01) (40, 4.53) (45, 5.02) (50, 4.68) (55, 3.65) (60, 3.72) (65, 3.92) (70, 6.89) (75, 13.37) (80, 16.74) (85, 11.67) (90, 8.94) (95, 8.82) (100, 9.47)
            };
        \addplot[smooth, color=teal, mark=+, mark size=2pt] coordinates {
                (0, 8.45) (5, 8.47) (10, 8.48) (15, 8.50) (20, 8.53) (25, 8.55) (30, 8.60) (35, 9.47) (40, 13.74) (45, 20.35) (50, 10.14) (55, 9.35) (60, 11.07) (65, 15.55) (70, 17.07) (75, 20.15) (80, 19.34) (85, 15.26) (90, 14.85) (95, 16.32) (100, 16.94)
            };
        \addplot[smooth, color=magenta, mark=diamond, mark size=2pt] coordinates {
                (0, 10.24) (5, 10.26) (10, 10.29) (15, 10.27) (20, 10.10) (25, 9.70) (30, 9.69) (35, 10.49) (40, 11.04) (45, 12.54) (50, 14.93) (55, 19.07) (60, 22.89) (65, 27.87) (70, 27.14) (75, 27.58) (80, 29.04) (85, 26.55) (90, 27.38) (95, 34.24) (100, 42.29)
            };
        \addplot[smooth, color=black, mark=o, mark size=2pt] coordinates {
                (40, 8.46) (45, 7.98) (50, 7.45) (55, 7.32) (60, 7.15) (65, 6.98) (70, 6.68) (75, 6.65) (80, 6.81) (85, 7.49) (90, 8.97) (95, 10.11) (100, 11.09)
            };
        \addplot[smooth, color=red, mark=asterisk, mark size=2pt] coordinates {
                (40, 1.92) (45, 1.91) (50, 1.94) (55, 1.99) (60, 2.08) (65, 2.18) (70, 2.53) (75, 2.92) (80, 3.30) (85, 4.18) (90, 5.71) (95, 6.83) (100, 7.78)
            };

        \nextgroupplot[
            ymin=-1, ymax=1, ytick={-1, -0.5, 0, 0.5, 1},
            xmin=0, xmax=100, xtick={0, 20, 40, 60, 80, 100},
            ylabel={\small $\Delta$LDR},
            y unit= dB
        ]
        \addplot[smooth, color=blue, mark=x, mark size=2pt] coordinates {
                (0, 0.005) (5, 0.006) (10, 0.006) (15, 0.006) (20, 0.005) (25, 0.004) (30, 0.001) (35, 0.026) (40, 0.140) (45, 0.189) (50, 0.202) (55, 0.092) (60, 0.054) (65, -0.013) (70, -0.257) (75, -0.522) (80, -0.546) (85, -0.228) (90, 0.038) (95, 0.145) (100, 0.190)
            };
        \addplot[smooth, color=teal, mark=+, mark size=2pt] coordinates {
                (0, 0.024) (5, 0.025) (10, 0.026) (15, 0.025) (20, 0.025) (25, 0.025) (30, 0.029) (35, 0.092) (40, 0.257) (45, 0.387) (50, 0.212) (55, 0.043) (60, -0.197) (65, -0.479) (70, -0.480) (75, -0.515) (80, -0.348) (85, 0.013) (90, 0.396) (95, 0.481) (100, 0.478)
            };
        \addplot[smooth, color=magenta, mark=diamond, mark size=2pt] coordinates {
                (0, 0.004) (5, 0.004) (10, 0.005) (15, 0.004) (20, -0.002) (25, -0.022) (30, -0.079) (35, -0.130) (40, -0.130) (45, -0.190) (50, -0.260) (55, -0.376) (60, -0.507) (65, -0.647) (70, -0.487) (75, -0.335) (80, -0.158) (85, 0.244) (90, 0.697) (95, 0.890) (100, 1.032)
            };
        \addplot[smooth, color=black, mark=o, mark size=2pt] coordinates {
                (40, -0.073) (45, -0.090) (50, -0.106) (55, -0.112) (60, -0.109) (65, -0.100) (70, -0.087) (75, -0.070) (80, -0.061) (85, -0.038) (90, -0.012) (95, 0.004) (100, 0.010)
            };
        \addplot[smooth, color=red, mark=asterisk, mark size=2pt] coordinates {
                (40, 0.001) (45, -0.023) (50, -0.048) (55, -0.061) (60, -0.064) (65, -0.062) (70, -0.065) (75, -0.058) (80, -0.057) (85, -0.040) (90, -0.016) (95, 0.002) (100, 0.010)
            };

        \nextgroupplot[ymin=-1, ymax=1, ytick={-1, -0.5, 0, 0.5, 1},
            xmin=0, xmax=100, xtick={0, 20, 40, 60, 80, 100}]
        \addplot[smooth, color=blue, mark=x, mark size=2pt] coordinates {
                (0, 0.007) (5, 0.007) (10, 0.007) (15, 0.006) (20, 0.005) (25, 0.004) (30, -0.004) (35, 0.001) (40, 0.111) (45, 0.173) (50, 0.182) (55, 0.046) (60, 0.033) (65, -0.033) (70, -0.292) (75, -0.574) (80, -0.644) (85, -0.312) (90, -0.051) (95, 0.031) (100, 0.055)
            };
        \addplot[smooth, color=teal, mark=+, mark size=2pt] coordinates {
                (0, 0.026) (5, 0.026) (10, 0.026) (15, 0.025) (20, 0.025) (25, 0.025) (30, 0.025) (35, 0.072) (40, 0.237) (45, 0.387) (50, 0.218) (55, 0.022) (60, -0.212) (65, -0.525) (70, -0.561) (75, -0.636) (80, -0.515) (85, -0.147) (90, 0.236) (95, 0.291) (100, 0.259)
            };
        \addplot[smooth, color=magenta, mark=diamond, mark size=2pt] coordinates {
                (0, 0.005) (5, 0.006) (10, 0.005) (15, 0.004) (20, -0.003) (25, -0.026) (30, -0.097) (35, -0.183) (40, -0.208) (45, -0.271) (50, -0.364) (55, -0.525) (60, -0.661) (65, -0.829) (70, -0.680) (75, -0.529) (80, -0.376) (85, 0.066) (90, 0.562) (95, 0.780) (100, 0.961)
            };
        \addplot[smooth, color=black, mark=o, mark size=2pt] coordinates {
                (40, -0.070) (45, -0.087) (50, -0.102) (55, -0.107) (60, -0.107) (65, -0.097) (70, -0.081) (75, -0.066) (80, -0.057) (85, -0.036) (90, -0.007) (95, 0.005) (100, 0.010)
            };
        \addplot[smooth, color=red, mark=asterisk, mark size=2pt] coordinates {
                (40, 0.005) (45, -0.021) (50, -0.045) (55, -0.056) (60, -0.065) (65, -0.060) (70, -0.060) (75, -0.056) (80, -0.054) (85, -0.039) (90, -0.010) (95, 0.005) (100, 0.013)
            };
    \end{groupplot}
\end{tikzpicture}

%% file: figures/interp.tex
\begin{tikzpicture}
    \begin{groupplot}[group style={
                    group size=1 by 2,
                    x descriptions at=edge bottom,
                    y descriptions at=edge left,
                },
            ybar=0pt,
            width=\columnwidth,
            height=0.5\columnwidth,
            xlabel={\small Peak Reduction},
            ylabel={\small ESR},
            y unit= \%,
            tick label style={font=\footnotesize},
            xtick={45, 55, 65, 75, 85, 95},
            ymajorgrids=true,
            grid style=dashed,
            ymax=11,
            ymin=6,
        ]
        \nextgroupplot[
            bar width=5pt,
            title={\normalsize Compressor},
        ]
        \addplot[line legend, sharp plot, blue, update limits=false] coordinates {(0, 7.49) (100, 7.49)} node [above] at (60, 7.49) {\footnotesize 7.49};
        \addplot[line legend, sharp plot, red, update limits=false] coordinates {(0, 7.56) (100, 7.56)} node [above] at (70, 7.56) {\footnotesize 7.56};
        \addplot[line legend, sharp plot, olive, update limits=false] coordinates {(0, 7.64) (100, 7.64)} node [above] at (80, 7.64) {\footnotesize 7.64};
        \addplot[fill=blue] coordinates {
                (45, 8.02) (55, 7.38) (65, 7.04) (75, 6.57) (85, 6.73) (95, 9.21)
            };
        \addplot[fill=red] coordinates {
                (45, 8.03) (55, 7.38) (65, 7.37) (75, 6.58) (85, 6.73) (95, 9.28)
            };
        \addplot[fill=olive] coordinates {
                (45, 8.04) (55, 7.39) (65, 7.30) (75, 6.64) (85, 6.73) (95, 9.73)
            };

        \nextgroupplot[
            bar width=5pt,
            title={\normalsize Limiter},
            legend style={legend columns=3, font=\footnotesize,
                    at={(0.5, -0.6)}, anchor=north},
        ]
        \addplot[line legend, sharp plot, blue, update limits=false] coordinates {(0, 7.76) (100, 7.76)} node [above] at (60, 7.76) {\footnotesize 7.76};
        \addplot[line legend, sharp plot, red, update limits=false] coordinates {(0, 7.84) (100, 7.84)} node [above] at (70, 7.84) {\footnotesize 7.84};
        \addplot[line legend, sharp plot, olive, update limits=false] coordinates {(0, 7.92) (100, 7.92)} node [above] at (80, 7.92) {\footnotesize 7.92};
        \addplot[fill=blue] coordinates {
                (45, 7.98) (55, 7.32) (65, 6.98) (75, 6.65) (85, 7.49) (95, 10.11)
            };
        \addlegendentry{Optimal}
        \addplot[fill=red] coordinates {
                (45, 8.01) (55, 7.32) (65, 7.32) (75, 6.67) (85, 7.52) (95, 10.19)
            };
        \addlegendentry{Linear}
        \addplot[fill=olive] coordinates {
                (45, 7.99) (55, 7.32) (65, 7.25) (75, 6.77) (85, 7.50) (95, 10.67)
            };
        \addlegendentry{Spline}

    \end{groupplot}
\end{tikzpicture}